\def\ba{\begin{eqnarray}}
\def\ea{\end{eqnarray}}
\def\lb{\label}
\def\nn{\nonumber \\}
\def\bi{\bibitem}

\def\g{\gamma}

\def\D{\Delta}

\def\pt{p_\perp}

\def\pr{\perp}
\def\sp{\;\;\;\;}

\documentclass[12pt]{article}
\usepackage{graphicx}
\usepackage{cite}

\usepackage{amssymb}

\usepackage{amsmath}
\usepackage[colorlinks=true,linktocpage=true,linkcolor=blue,citecolor=blue]{hyperref}

\begin{document}

\title{Short-range two-particle  correlations from statistical clusters}

\author{A.Bialas$^a$ and A.Bzdak$^b$
\vspace{2mm}
\\ $^a$M.Smoluchowski Institute of Physics\\Jagellonian University\thanks{Address: Lojasiewicza 11, 30-348 Krakow, Poland; e-mail: bialas@th.if.uj.edu.pl}
\\ $^b$AGH University of Science and Technology\\Faculty of Physics and Applied Computer Science\thanks{Address: 
al. Mickiewicza 30, 30-059 Krakow, Poland; e-mail: bzdak@fis.agh.edu.pl}
}
\maketitle



\begin{abstract}
The two-particle short-range correlation functions in rapidity, azimuthal angle and transverse momentum, following from the decay of statistical clusters are evaluated and discussed.
\end{abstract}

\section {Introduction}
Studies of  correlations between particles produced in high-energy collisions is a well-known method to investigate the dynamics of the production  process\cite{cor}. They are conveniently divided into  "short-range"  when the momenta of the studied particles are close to each other and "long-range", extending over  large distances in momentum space.

Already in early seventies, studies of the short-range correlations in rapidity  led to the discovery that particle production proceeds through production of "clusters" \cite{abbb}. This was later confirmed by more detailed studies in other variables \cite{cor}, although the very nature of these clusters is unclear even now.  The problem is of importance because  it touches the mechanism of hadronisation, i.e., transition from the parton system, created at the early stage of the collision, into the produced hadrons. This  apparently non-perturbative transition, essential  to derive the structure of the produced parton system from the observed hadrons, cannot be easily treated by theory.  Thus a phenomenological analysis is needed.

An interesting  approach to this problem was formulated in the  statistical cluster model  \cite{becattini1,bclus} which assumes that the transition from the early state of the process of particle production, dominated by parton interactions, proceeds through an intermediate stage of clusters emitting (isotropically) the final hadrons according to the rules of statistical physics\footnote{ 
This is an attractive modification of the standard statistical model,  because it allows to explain the observed anisotropy of the  momentum spectra of particles produced in high-energy collisions (which is a difficulty for the  statistical model), while keeping, at the same time, its successes in description of  particle abundances \cite{and,becattini2}.}.

The decay distribution of such a statistical cluster at rest is taken in the form of the Boltzmann distribution, which for a cluster moving with the four-velocity $u^\mu$ becomes  
\ba
dN_1(p;u)\sim e^{-\beta p_\mu u^\mu }d^2\pt dy ,  \lb{single}  \lb{mc}
\ea 
where $\beta=1/T$ is the inverse cluster temperature and $\pt$ and $y$ are the transverse momentum and rapidity of the final particle.

Although the model was originally constructed for description of the "soft" processes (involving only small transverse momenta),  it remains an interesting question to what extent it is also applicable to semi-hard and perhaps even hard collisions. Indeed, the patron-hadron transition being a soft process, happening at the very end of the parton cascade,  may very well be universal, i.e.,  (quasi)independent of the mechanism of parton production. If this is actually the case, the statistical clusters should be visible also at higher transverse momenta and perhaps even in all processes of particle production at high energies. This attractive possibility was recently supported by the evaluation \cite{abts} of the transverse momentum spectrum  of the produced charged hadrons. It turned out that if the distribution of the cluster transverse Lorentz factor  $\g_\pr$ ($ \g_\pr^2 =1+u_\pr^2$, where $u_\pr$ is the transverse component of the cluster four-velocity)  follows a simple power law $\sim \g_\pr^{-\kappa}$ then, surprisingly enough, the transverse momentum distribution of the emitted particles 
 \ba
\frac{d N_1(p_\pr)} {dp_\pr^2}\sim \int \frac{d\g_\pr}{\g_\pr^{\kappa}} K_0(\beta\g_\pr m_\pr)I_0 (\beta u_\pr p_\pr)  \lb{sg}
 \ea
closely resembles the Tsallis formula \cite{tsallis}  which, as is well-known \cite{ww1,cl2,Wong:2015mba}, closely resembles the shape of the  data \cite{phenix1,alice1,atlas1,cms1}. See \cite{abts} for more details and \cite{Wilk:2015yha} for further discussion. 

This apparent success  of the concept of the statistical cluster invites  one to study its other consequences, particularly those which may provide  more demanding tests of the idea. Following this route, in the present paper we discuss the two-particle correlations and show that, indeed, they give strong constraints on the model and, eventually,  can be even used  to pin down possible inter-cluster correlations\footnote{To our knowledge, the first application of the statistical model for description of cluster decays was proposed by Hayot, Henyey and Le Bellac \cite{abbb}.}.

\begin{figure}[t]
\begin{center}
\includegraphics[scale=0.5]{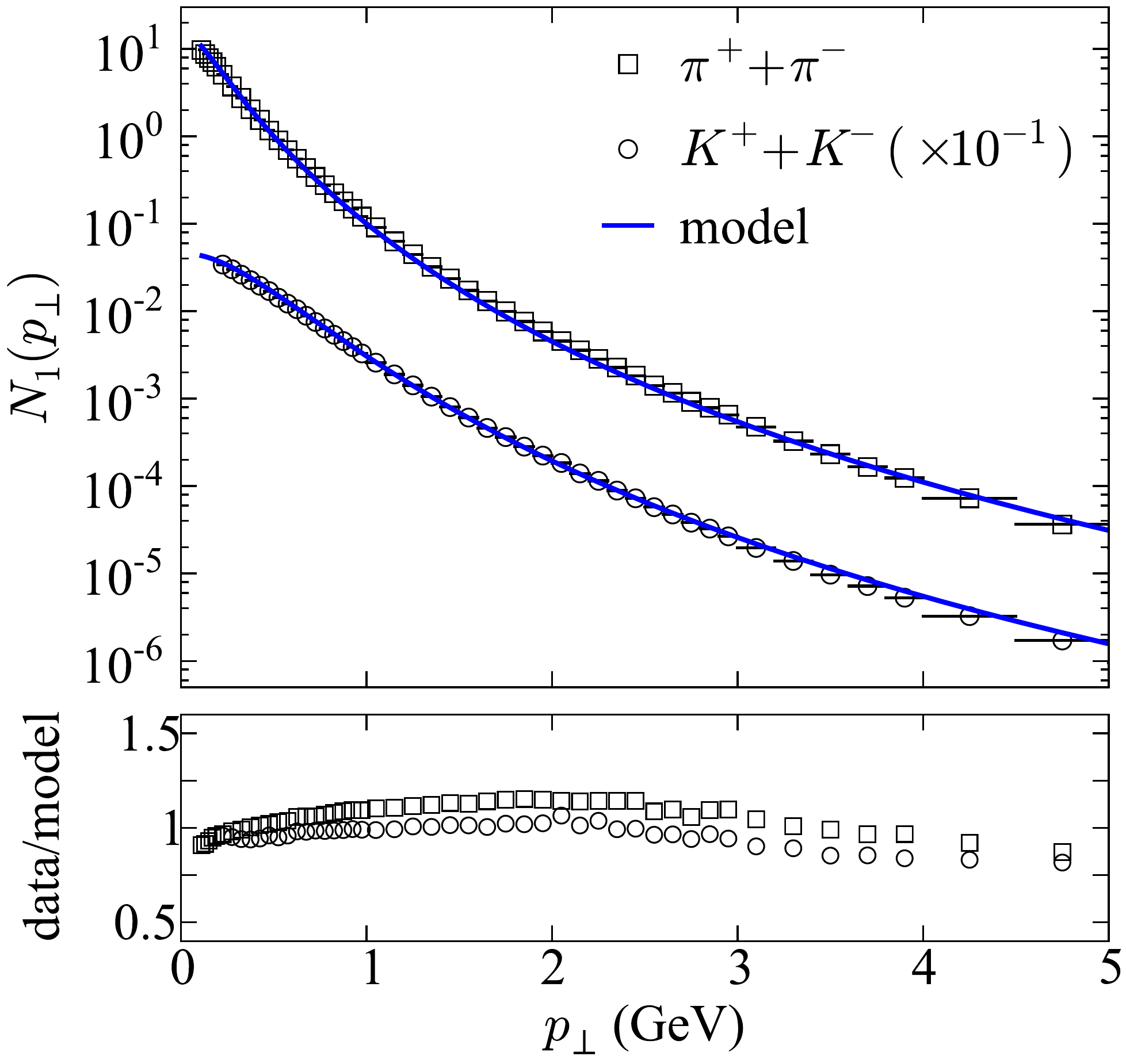}
\end{center}
\par
\vspace{-5mm}
\caption{The single particle distributions for pions and kaons as measured by ALICE in p+p collisions at $\sqrt{s}=2.76$ TeV, compared to the statistical cluster model with $T=140$ MeV and $\kappa=5$. The ratios data/model are shown at the bottom.}
\label{fig:intro}
\end{figure}

To determine $T$ and $\kappa$, the only free parameters of the model, we have  fitted the cluster formula (\ref{sg}) to  the transverse momentum distribution of pions and kaons produced in 2.76 TeV proton-proton (p+p) collisions \cite{Abelev:2014laa}. The fit gives $\kappa=5$ and the cluster temperature $T=140$ MeV. The result is shown in  Fig. \ref{fig:intro}, where one sees that the model reproduces the data with better than $20$\% accuracy, which is good enough for our purpose. 
It is also remarkable that the same power-law distribution (without change of normalization) describes well both pion  and kaon distributions. We also checked that the model describes  the charged particle spectra up to $p_{\pr}=200$ GeV in p+p at $\sqrt{s}=7$ TeV \cite{Chatrchyan:2011av}.

In the next section the general formula for the two-particle correlations in the statistical cluster model is written down, and correlations in rapidity, azimuthal angle and in transverse momentum are derived. Our results are described in Section 3. Summary and comments are given in the last section.

\section{Two-particle correlations}

The  two-particle  distribution is the sum of the contribution from one cluster and that from two different clusters $dN_2=dN_2^{(1c)}+dN_2^{(2c)}$. Ignoring correlations in cluster decay (see section 4 for further discussion) we have
\ba
dN_2^{(1c)}(p_1,p_2) =\int du W(u) dN_1(p_1;u)dN_1(p_2;u)   \lb{dn}
\ea 
and
\ba
dN_2^{(2c)} =\int du_1 \int du_2 W(u_1,u_2) dN_1(p_1;u_1)dN_1(p_2;u_2) ,
\ea
where $W(u)$ is the distribution of (four)velocity of a cluster and $W(u_1,u_2)$ is the  corresponding distribution of two clusters\footnote{$\int W(u)du=\langle N\rangle $; $\int du_1du_2W(u_1u_2)=\langle N(N-1)\rangle$ where $N$ denotes number of clusters.}. {For the two particle correlation function 
\ba 
 C(p_{1},p_{2})d^2 p_{1\pr} dy_1d^2 p_{2\pr} dy_2 \equiv dN_2(p_{1},p_{2})-dN_1(p_{1})dN_1(p_{2}),
 \ea
 with $ dN_1(p) =\int du W(u) dN_1(p,u)$, we thus  have
 \ba
 C(p_{1},p_{2})d^2 p_{1\pr} dy_1d^2 p_{2\pr} dy_2=\nn=
 dN_2^{(1c)}(p_{1},p_{2})+\int du_1 \int du_2 C_u(u_1,u_2) dN_1(p_{1};u_1) dN_1(p_{2};u_2).  \lb{clf}
 \ea
where 
 \ba
 C_u(u_1,u_2)=W(u_1,u_2)- W(u_1) W(u_2)
 \ea
is  the two-cluster correlation function.

If clusters are independent, i.e., $C_u(u_1,u_2)=0$, we have}
 \ba
 C(p_{1},p_{2})d^2 p_{1\pr} dy_1d^2 p_{2\pr} dy_2=dN_2^{(1c)}(p_{1},p_{2}). \lb{corr}
 \ea 
Consider a cluster\footnote{Henceforth we assume that clusters are uncorrelated.} at rapidity $Y$ moving in the transverse direction with the velocity $v_\perp$. 
We have
\ba
u_{0}=\g_{\pr}\cosh Y;\sp u_{z}=\g_{\pr}\sinh Y;\sp u_{\pr}=\g v_{\pr};\sp
v_{z}=\tanh Y.
\ea
and the formula (\ref{mc}) becomes
\ba
dN_1(p,u)=e^{-\beta \g_\pr m_\pr \cosh(y-Y)+\beta p_\pr u_\pr\cos(\phi_u - \phi)} ,
\ea
where $\phi_u$ and $\phi$ are the azimuthal angles of the cluster and of the produced particle, respectively.

Following \cite{abts} we take
\ba
W(u)du \sim \g_\pr ^{-\kappa} d\g_\pr d\phi_u G(Y) dY , \lb{gv}
\ea
where $G(Y)$ is the distribution of clusters in rapidity.\footnote{We note that our results do not depend on the specific shape of $G(Y)$.}

\subsection {Correlation in rapidity and azimuthal angle}

We start with  the correlations  in rapidity and azimuthal angle.
Using the formulae of the previous section we have
\ba
dN_2^{(1c)}= dy_1  d^2p_{1\pr} dy_2 d^2p_{2\pr}  \int \g_\pr ^{-\kappa} d\g_\pr \int d\phi_u dY G(Y)\nn
e^{-\beta \g_\pr [m_{1\pr} \cosh(y_1-Y)+m_{2\pr} \cosh(y_2-Y)]}e^{\beta u_\pr[p_{1\pr} \cos(\phi_u-\phi_1)+p_{2\pr} \cos(\phi_u-\phi_2)]}.
\ea
To obtain the distributions of $y_1-y_2\equiv \Delta y$ and $\phi_1-\phi_2\equiv \Delta \phi$ at fixed $p_{1\pr}$ and $p_{2\pr}$ we integrate over $\phi_u$, $y_+=y_1+y_2$, $\phi_+=\phi_1+\phi_2$ and $Y$. The result is 
\ba
C(\D y, \D\phi)\sim \int \frac{ d\g_\pr}{\g_\pr ^{\kappa}} K_0\left[\beta \g_\pr D_m(\D y)\right]  I_0[\beta u_\pr D_p(\D\phi)] , 
\ea
where $I_0$ and $K_0$ are the modified Bessel functions of the first and second kind, respectively, and
\ba
D_m(\D y) &=& \sqrt{m_{1\pr}^2+m_{2\pr}^2+2m_{1\pr}m_{2\pr}\cosh(\D y)} , \nn
D_p(\D\phi) &=& \sqrt{p_{1\pr}^2+p_{2\pr}^2+2p_{1\pr}p_{2\pr}\cos(\D\phi)}.
\ea
Correlations in $\D \phi$ at fixed $p_{1\pr}$ and $p_{2\pr}$ can be obtained by integrating independently $y_1$ and $y_2$ with the result\footnote{{We skip the factors [$2\pi-\D \phi$] (for $\D \phi <\pi$) and  $[\D \phi]$ (for $\D \phi >\pi$) since they are canceled when $C(\D \phi)$ is divided by the distribution of  mixed events.}}
\ba
C(\D \phi)\sim \int \frac{ d\g_\pr}{\g_\pr ^{\kappa}}  K_0[\beta \g_\pr m_{1\pr}] K_0[\beta \g_\pr m_{2\pr}]   I_0\left[\beta u_\pr D_p(\D\phi)\right] . \lb{cfi}
\ea
For fixed $\D y$ we have, similarly, 
\ba
C(\D y)\sim \int \frac{ d\g_\pr}{\g_\pr ^{\kappa}}  K_0\left[\beta \g_\pr D_m(\D y)\right]I_0[\beta u_\pr p_{1\pr}]I_0[\beta u_\pr p_{2\pr}] .  \lb{cy}
\ea

\subsection {Correlation in transverse momentum}

It is also interesting to consider  the distribution of moduli of transverse momenta $[p_{1\pr}, p_{2\pr}]$. Integrating $C(p_{1},p_{2})$  over $y_1,y_2,\phi_1,\phi_2, Y, \phi_u$, one obtains 
\ba
C(p_{1\pr},p_{2\pr}) \sim \int \frac{d\g_{\pr}}{\g_{\pr}^\kappa}K_0[\beta \g_\pr m_{1\pr}]K_0[\beta \g_\pr m_{2\pr}]
I_0[\beta u_\pr p_{1\pr}]I_0[\beta u_\pr p_{2\pr}].  \lb{corpt}
\ea

\section {Results}

As  explained in the introduction, the two parameters of the model: the temperature in the cluster decay $T= 140$ MeV  and the power $\kappa=5$ were determined  from the fit to the pion and kaon single-particle transverse momentum distributions measured by ALICE \cite{Abelev:2014laa}.

\begin{figure}[t]
\begin{center}
\includegraphics[scale=0.5]{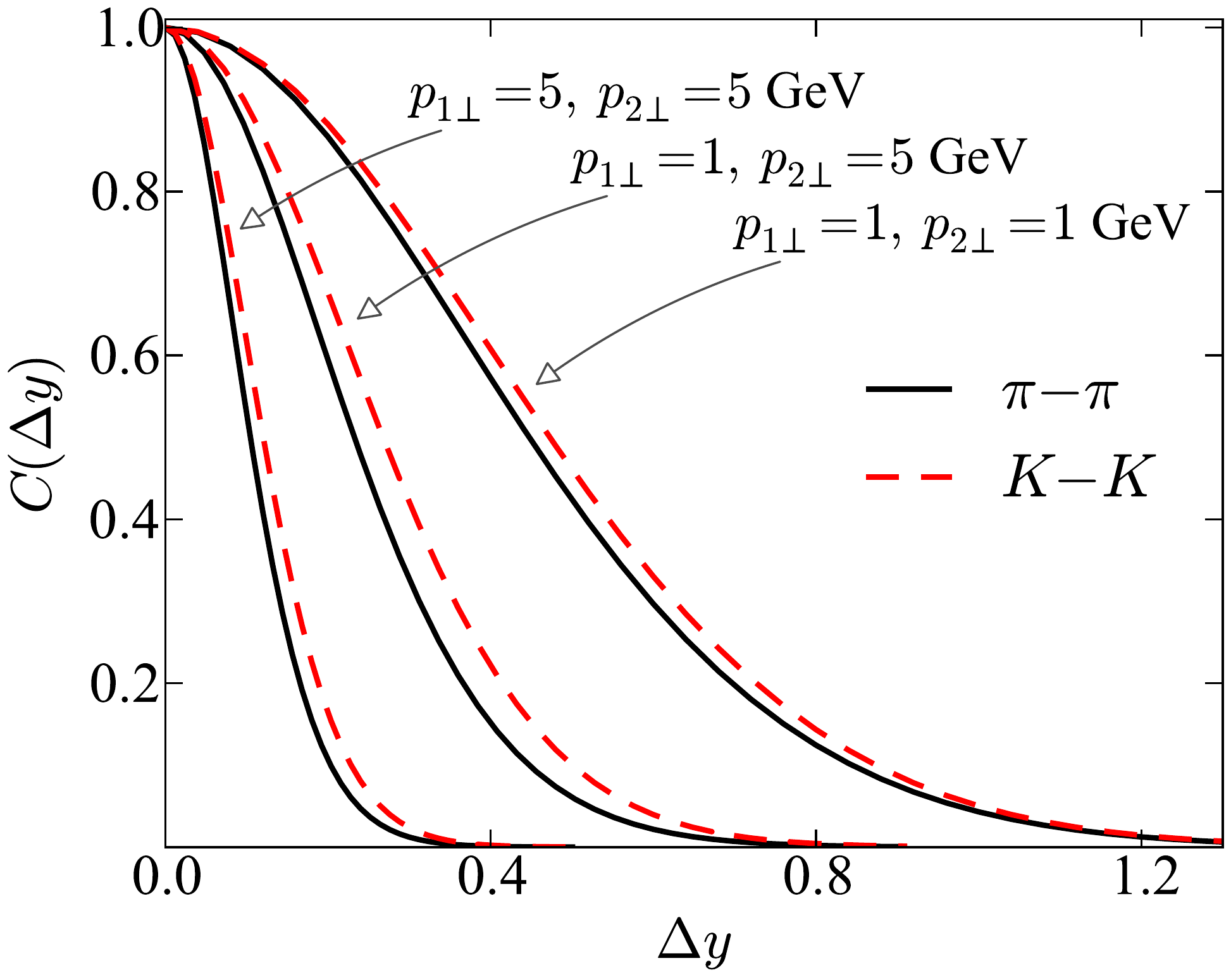}
\end{center}
\par
\vspace{-5mm}
\caption{The two-particle correlation function (\ref{cy}) from a statistical cluster for pairs of pions and kaons with various values of the transverse momenta, plotted vs. $\Delta y=y_1-y_2$, the rapidity separation between the two particles.  $T=140$ MeV and $\kappa=5$. $C$ is scaled to $1$ at $\Delta y=0$.}
\label{fig:1}
\end{figure}

\begin{figure}[t]
\begin{center}
\includegraphics[scale=0.5]{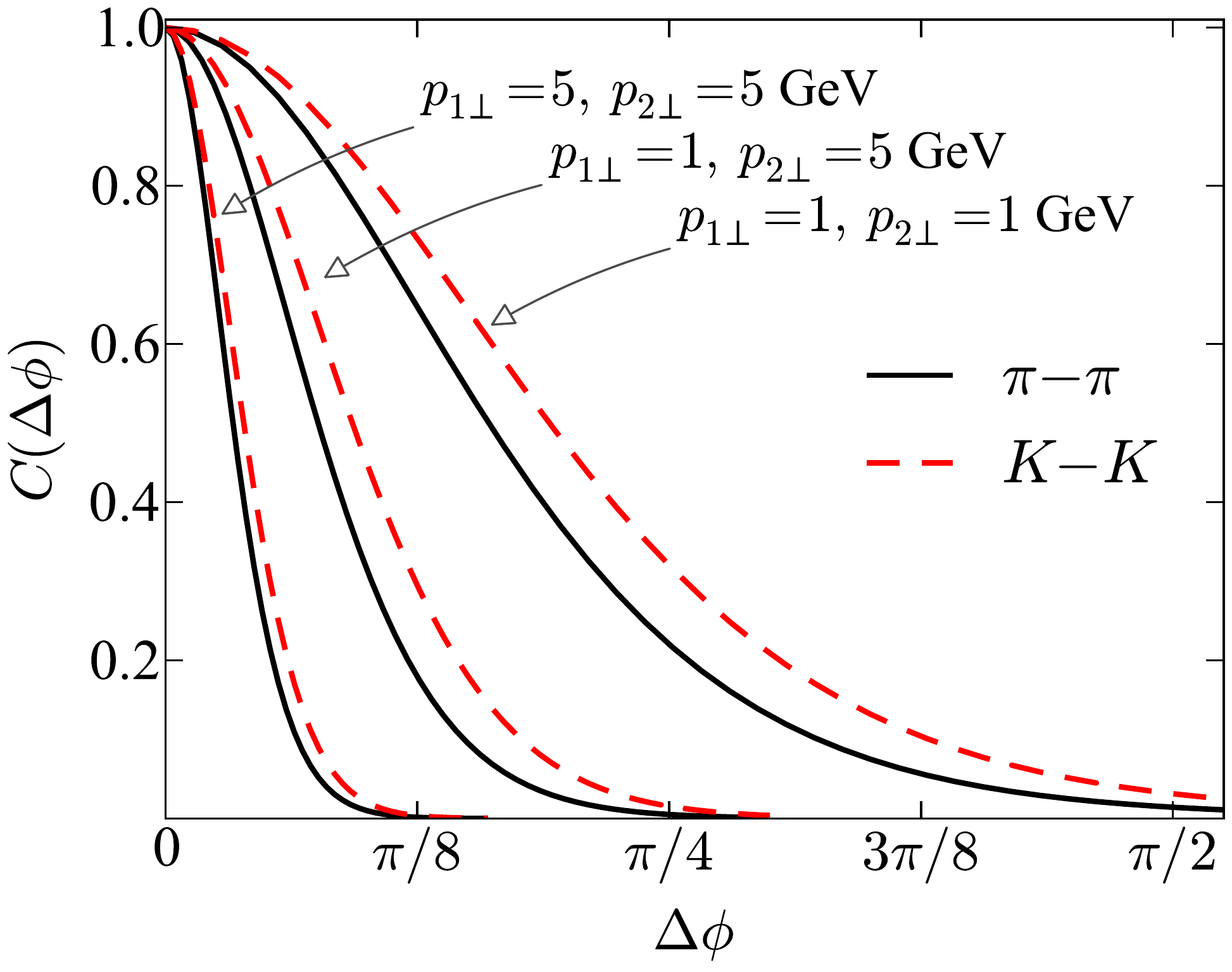}
\end{center}
\par
\vspace{-5mm}
\caption{The two-particle correlation function (\ref{cfi}) from a statistical cluster for pairs of pions and kaons and with various values of the transverse momenta, plotted vs. $\Delta\phi=\phi_1 - \phi_2$,  the relative azimuthal angle between the two particles. In this calculation $T=140$ MeV and $\kappa=5$. $C$ is scaled to $1$ at $\Delta\phi=0$.}
\label{fig:2}
\end{figure}

The correlation functions in azimuthal angle and  rapidity, given by (\ref{cfi}) and (\ref{cy}), are shown in Figs. \ref{fig:1}, \ref{fig:2}. 
In Fig. \ref{fig:1} the correlation function  $C(\Delta y; p_{1\pr},p_{2\pr})$ (normalised  to $1$ at $\Delta y=0$) is plotted {vs}  $\Delta y=|y_1-y_2|$, for pairs of pions and kaons, at various values of the transverse momenta.  One sees that $C$ gets narrower with increasing  $p_{1\pr}$ and $p_{2\pr}$ and there is also some  mass dependence. The correlation function $C(\Delta \phi; p_{1\pr},p_{2\pr})$ is plotted in Fig.  \ref{fig:2}. Similar features are also seen,  except that the dependence on particle mass is more pronounced. 

Numerical calculations show that for sufficiently high transverse momenta (above  $\sim$ 2 GeV) and vanishing particle masses, the two-particle correlation functions in rapidity and in azimuthal angle,  can be approximated by Gaussians with the width squared proportional to $T^2$ and inversely proportional to the product $p_{1\pr}p_{2\pr}$. The proportionality factor is close to $2\kappa$.
 
\begin{figure}[t]
\begin{center}
\includegraphics[scale=0.4]{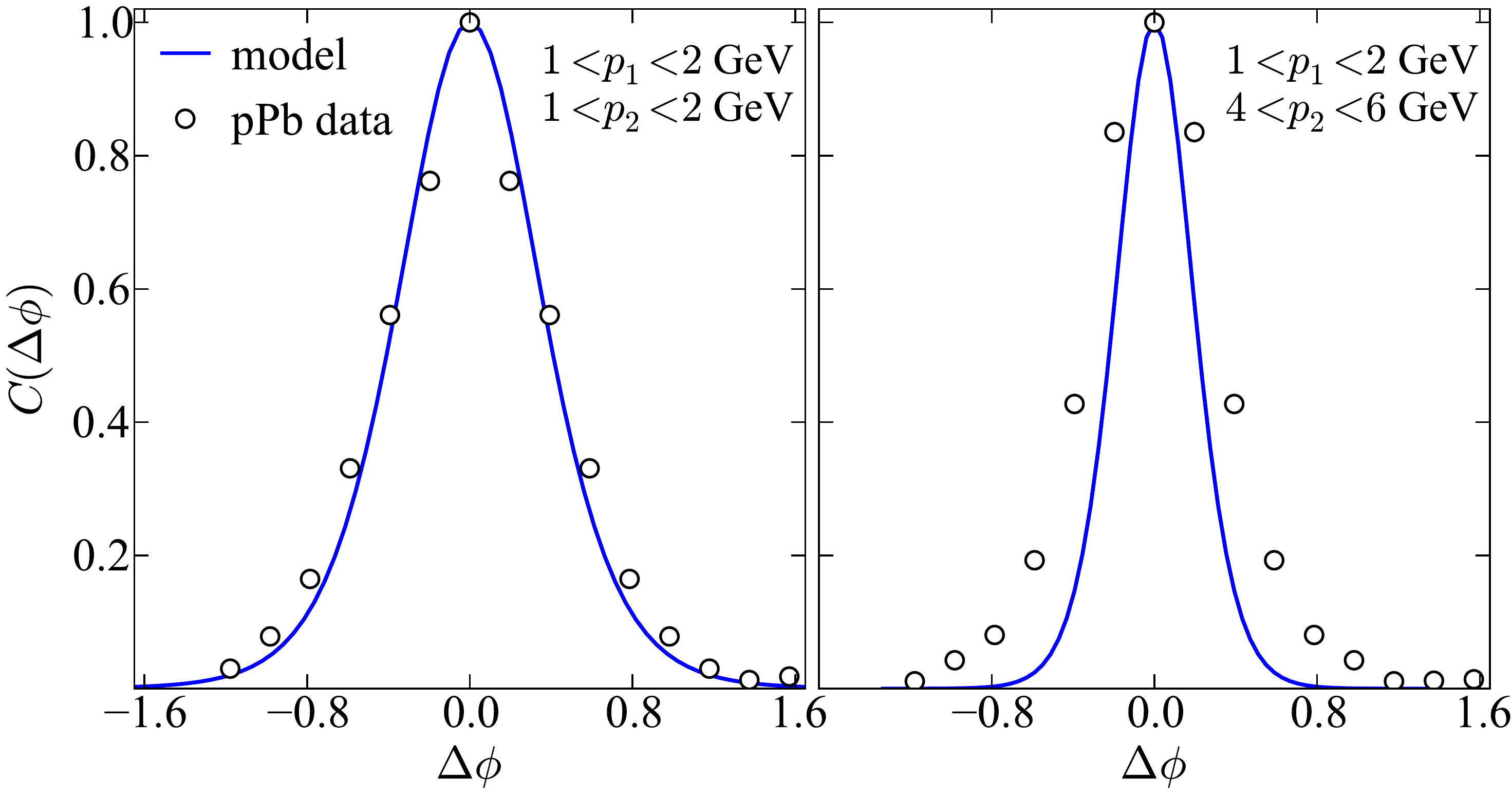}
\end{center}
\par
\vspace{-5mm}
\caption{The two-particle correlation function from a statistical cluster for low (left) and high (right) transverse momenta compared with the CMS data \cite{Chatrchyan:2013nka} on the short-range azimuthal correlation function  measured in p+Pb collisions at $\sqrt{s}=5.02$ TeV.  $T=140$ MeV, $\kappa=5$. All functions are scaled to $1$ at $\Delta\phi=0$.}
\label{fig:cms}
\end{figure}

Recently the CMS Collaboration  published \cite{Chatrchyan:2013nka} extensive studies of the two-particle azimuthal correlation functions in p+Pb collisions at $\sqrt{s}=5.02$ TeV. In Fig. \ref{fig:cms} they are compared with the predictions of the statistical cluster model. One sees that the data are reasonably close to the model predictions at transverse momenta in the region of $1-2$ GeV. At higher transverse momenta the model gives correlation functions which seem somewhat too narrow.\footnote{The published CMS data \cite{Chatrchyan:2013nka} are modified by other physical effects, e.g., flow in p+Pb, the back-to-back peak in $\Delta \phi$ (which is not considered in the present paper), and by the procedure of the background removal.} 

\begin{figure}[t]
\begin{center}
\includegraphics[scale=0.5]{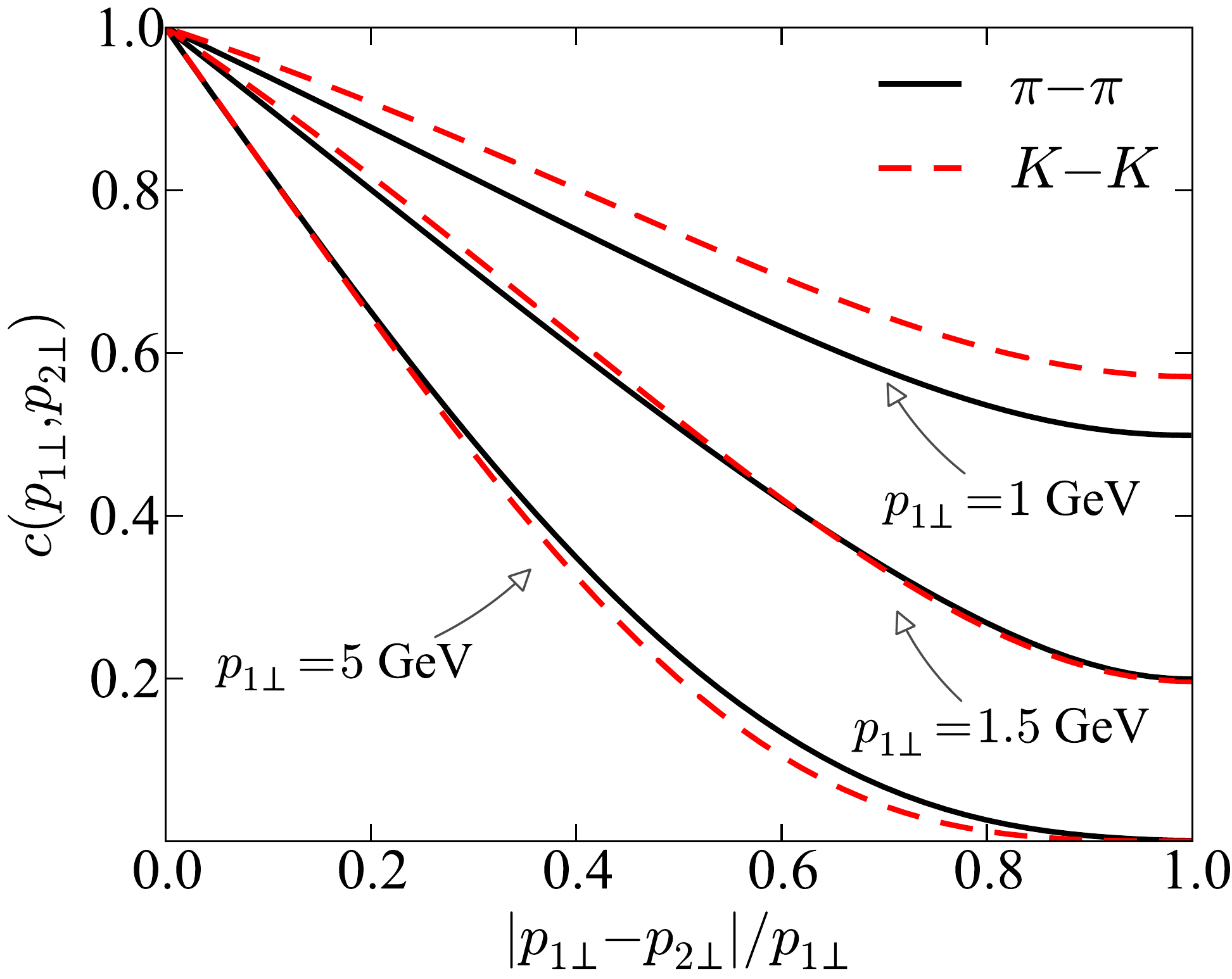}
\end{center}
\par
\vspace{-5mm}
\caption{The transverse momentum correlation function (\ref{corptNN}) from a statistical cluster for pairs of pions and kaons vs. 
$(p_{1\pr}-p_{2\pr})/p_{1\pr}$, for various values of $p_{1\pr}$. Here $T=140$ MeV and $\kappa=5$. $c$ is scaled to $1$ at $p_{1\pr}-p_{2\pr}=0$.}
\label{fig:pt1}
\end{figure}

\begin{figure}[t]
\begin{center}
\includegraphics[scale=0.5]{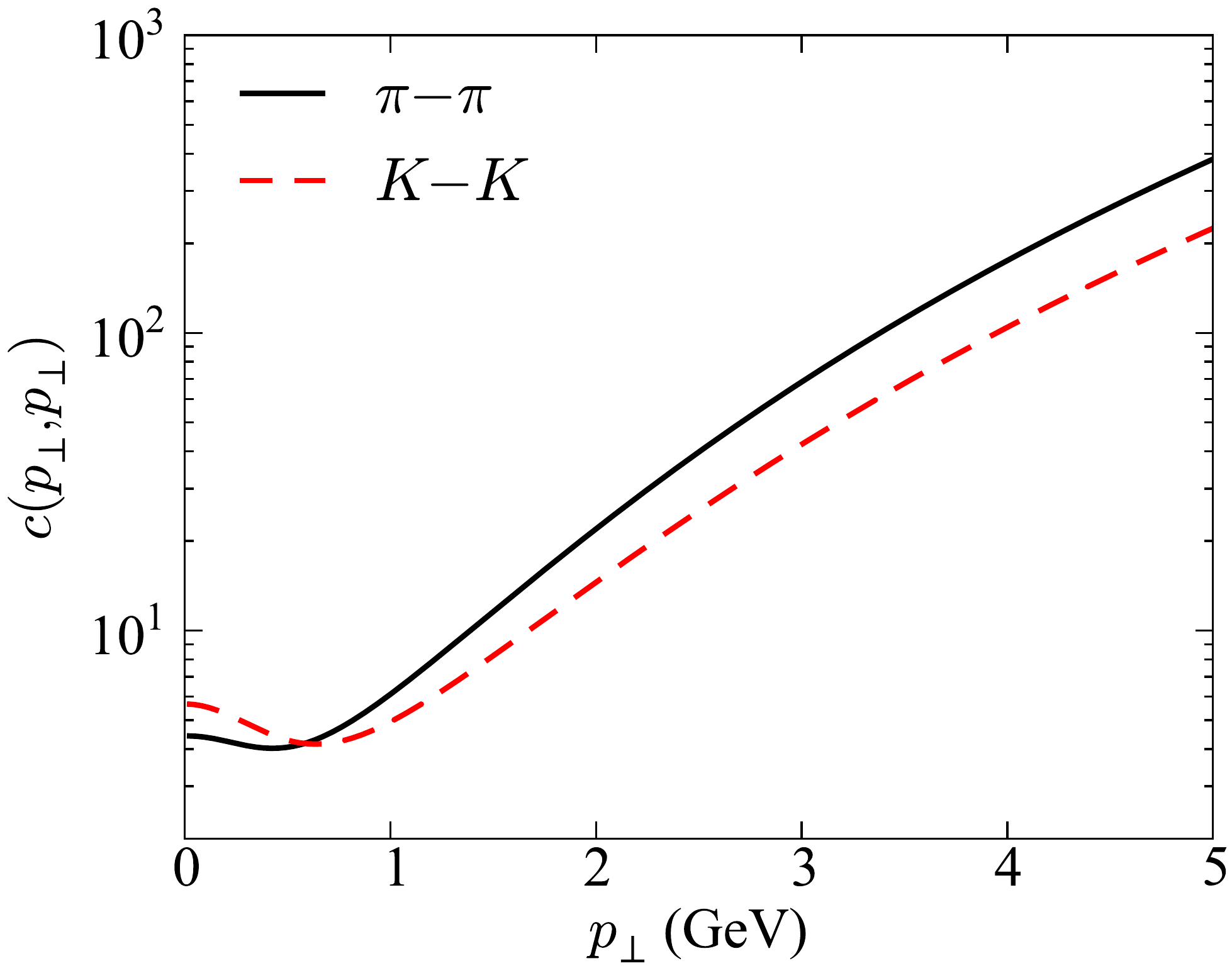}
\end{center}
\par
\vspace{-5mm}
\caption{The transverse momentum correlation function (\ref{corptNN}) from a statistical cluster for pairs of pions and kaons vs. $p_{\pr}=p_{1\pr}=p_{2\pr}$. $T=140$ MeV and $\kappa=5$.}
\label{fig:pt2}
\end{figure}

This result agrees with the idea that the model is applicable only at relatively small transverse momenta. On the other hand it may perhaps also indicate the presence of cluster-cluster correlations (described by the second term in the R.H.S. of (\ref{clf})) at transverse momenta above $2$ GeV. Possible resolution of  this dilemma would require more sophisticated studies and goes beyond the scope of this paper.

It would be also interesting to study balance functions \cite{bfdef,bfun,bfpr}. This is not possible at the present stage of the model since it requires additional information on the distribution of the cluster charges.  

The transverse momentum correlation (\ref{corpt}), divided by the product of the two single-particle distributions\footnote{It is convenient to use this definition of $c(p_{1\pr},p_{2\pr})$ since it is proportional to the number of pairs divided by the number of pairs in mixed events.}, 
\ba
	c(p_{1\pr},p_{2\pr})=\frac{C(p_{1\pr},p_{2\pr})}{N_1(p_{1\pr})N_1(p_{2\pr})} \lb{corptNN}
\ea
is shown in Figs. \ref{fig:pt1} and \ref{fig:pt2}. $c(p_{1\pr},p_{2\pr})$, normalized to 1 at $p_{1\pr}=p_{2\pr}$, is plotted in Fig. \ref{fig:pt1} vs. the ratio $|p_{1\pr}-p_{2\pr}|/p_{1\pr}$ (with $p_{2\pr} \le p_{1\pr}$) for various values of 
$p_{1\pr}=1, 1.5, 5$ GeV.\footnote{We checked that above  $p_{1\pr} =5$ GeV the curves practically do not change any more.}
In Fig. \ref{fig:pt2} the value of $c$ at $p_{\pr}=p_{1\pr}=p_{2\pr}$ is plotted vs. $p_{\pr}$. One sees a rather fast increase of $c$ with increasing $p_{\pr}$.

\section {Summary and comments}

In summary, we have  constructed the two-particle correlation functions induced by the decay of a statistical cluster. The explicit formulae were given for correlations in rapidity, azimuthal angle and in transverse momentum. Using the parameters of the model determined from the fit to the single-particle distributions ($\kappa=5,\; T=140$ MeV) the correlation functions were evaluated. 
Qualitative comparison with the  CMS data on azimuthal correlations in p+Pb collisions at $\sqrt{s}=5.02$ TeV shows that the model works well at transverse momenta around $1-2$ GeV. For larger transverse momenta the evaluated correlation function looks somewhat too narrow, possibly  indicating presence of additional inter-cluster correlations. 


Several comments are in order.

(i)  It should be emphasised that the present calculation follows directly from the statistical cluster model and thus  contains no free parameters: the value of the freeze-out temperature ($T\sim$ 140 MeV) and  the parameter $\kappa=5$  were  determined  from the single-particle transverse momentum distribution of pions and kaons. It seems remarkable that 
spectra of both  pions and kaons can be described simultaneously with exactly the same cluster distribution.

 
(ii) As already explained in the Introduction, we hypothesize that at the final stage of the production process, the statistical clusters are formed and decay into observed particles. At high transverse momenta,  jet physics is expected to induce correlations between clusters and thus additional correlations between produced particles. Detailed experimental investigation of this region could therefore verify universality of  the cluster hypothesis and may also give useful information on the structure of jets.

(iii) It would be also most interesting to measure and compare the short-range correlation functions in p+p and e$^{+}$e$^{-}$ collisions in order to test universality of the statistical cluster picture of particle production. Also measurement of correlations for various pairs of particles can be very useful in this respect.

(iv) In our calculations we have ignored the correlations which may appear in the cluster decay.
The statistical clusters are rather special objects and their physical interpretation, and consequently the nature of their internal correlations, is (in our opinion) not clear. For example, it is not obvious if clusters, representing multi-particles states, have a well-determined mass. The detailed comparison of the model with data in p+p or e$^{+}$e$^{-}$ collisions should shed more light on these questions.

(v) In the present paper we have not discussed the baryon production, as it is not clear if at LHC energies  the statistical model can describe correctly the baryon multiplicities. Within the statistical cluster model one may overcome this difficulty, e.g., by postulating that the clusters emitting baryons are of different nature than those producing mesons only. This problem is under investigation.

\bigskip
\noindent
{\bf Acknowledgements}
\vspace{1mm}

We thank Kacper Zalewski for very useful discussions. This investigation was supported by the Ministry of Science and Higher Education (MNiSW), by founding from the Foundation for Polish Science, and by the National Science Centre (Narodowe Centrum Nauki), Grant Nos. DEC-2013/09/B/ST2/00497 and DEC-2014/15/B/ST2/00175.

\end{document}